# Association Between Gold Grain Orientation And Its Periodic Steps Formed At The Gold/Substrate Interface

Linfeng Chen,[†] Maria Koifman Khristosov,[†,‡] Cecile Saguy,[§] Alex Katsman,[†] and Boaz Pokroy*[,†,‡]

[†]Department of Materials Science and Engineering, Technion – Israel Institute of Technology, Haifa 32000, Israel

[‡]The Russell Berrie Nanotechnology Institute, Technion—Israel Institute of Technology, 3200 Haifa, Israel

[§]Solid State Institute, Technion – Israel Institute of Technology, 32000 Haifa, Israel.

*Boaz Pokroy

e-mail: bpokroy@tx.technion.ac.il; Home page: http://pokroylab.net.technion.ac.il

**ABSTRACT:** Nanoscale step structures have attracted recent interest owing to their importance in both fundamental and applied research, for example in adsorption, in catalysis, and in directing nanowire growth. Here, we used a template-stripping method to obtain vicinal-like surface structures on grains of polycrystalline gold and investigated the effect of annealing temperature on the formation of these surfaces. Our results uncovered a correlation between the grain orientation angle (GOA) and the step periodicity and crystallographic direction **on identical grains**. The GOA **was** measured by determining the electron backscatter **diffraction with respect** to **the sample's normal direction**. Using scanning tunneling microscopy to examine identical grains, we found that their step periodicity decreases with increasing GOA. These **results provide further** understanding of the formation of periodically atomic gold steps at the gold/substrate **interface, and thus might have promising potential for the directing growth** of nanowires in microelectronics.

**TOC**

**GRAPHIC**

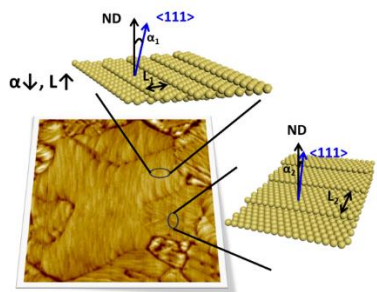

**KEYWORDS** EBSD, STM, step structures, template stripping, *in situ*.

Nanometer-scale structures of less than 10 nm on a solid surface are **attracting growing** interest in the fields of nanoscience and nanotechnology, motivated by their potential applications in electronics,[1-4] magnetic storage devices,[5,6] optics[7,8] and molecular switches.[9,10] **Fabrication** of such nanoscale structures in a controlled way remains challenging, and **is not feasible by** conventional top-down techniques such as lithography, writing or stamping, which are usually capable of creating features as low as the sub-100-nm range.[11] **Moreover, architectural fabrication** at a length scale of **less than 10 nm for molecular device fabrications[4,13]** using a **bottom-up approach[11,12] has low and insufficient control** for guiding **the on-place** assembly of molecules, which is especially important for electronics.

In recent decades, **vicinal surface templates were developed** to combine **the advantages** of the top-down and bottom-up **strategies, respectively their ease of fabrication and** exquisite control over their mesoscale organization.[14] **Vicinal surfaces, which are characterized by a periodic succession of nanoterraces and atomic steps, are** of particular interest and have been widely investigated for catalysis,[15] polymerization,[16] single molecular assemblage,[17,18] guiding the growth of nanowires,[19,20] and assembly of ordered Co nanodot arrays[21] and C60 nanochains.[22] **One way** to produce vicinal surfaces is by miscutting a single crystal at a small angle with respect **to its low-index plane**, followed by mechanical polishing or ion bombardment and high-temperature annealing. However, this process is costly, tedious, and limited to single crystals.[23, 24]

We recently developed a new way to create polycrystalline metal surfaces having similar structures to those of vicinal surfaces,[25] by **utilizing the simple method of template stripping (TS)**.[26] We showed that following evaporation of thin metallic film (e.g. gold), then annealing and finally TS, an ultra-smooth surface is obtained, with each **grain** having a vicinal-like surface structure. We also showed that **the periodicity** of the vicinal steps can be controlled by changing the **substrate used for** TS. Nevertheless, additional details of the process and of the parameters affecting step periodicity, in particular the annealing temperature needed to create the step structure, remain unknown. Moreover, the relationship between the step periodicity and the grains' orientation angle is still not clear. In the present work we **investigated _in situ_** the effect of annealing temperature on the creation of vicinal-like surfaces. We also investigated the correlation **between the normal direction of the grain orientation angle (GOA) in our sample** and the corresponding step periodicity of the same grain.

Scanning tunneling microscopy (STM) showed that the inner surfaces were microscopically flat **for both samples (Au-ref, Figure 1b, and Au-T400t1h, Figure 1d)**. Further magnification of the nanostructure (Figure 1c,e) revealed a remarkable difference between the samples upon annealing: in the Au-T400t1h **grains** (Figure 1e,g, atomically self-ordered vicinal-like structures were formed, whereas Au-ref was found to be an atomically flat nanogranular structure (Figure 1c,f). Consistently with our previous results,[27] the step structure **in the given grain in Au-T400t1h** was well ordered **and all steps had** the same orientation. Step height was in the range of **0.3 nm, with only single atomic steps observed** between terraces (Figure 1e, inset). Figures 1f and 1g depict **the 3D representations corresponding to** Figures 1c and e. To verify that the step structure formation is attributable to the thermal treatment and not to contamination, both samples (with and without annealing) were examined by X-ray photoelectron spectroscopy (XPS). The **results confirmed** that the inner surfaces of the gold **films, both before** and after 1 h of annealing at 400°C, were of pure gold (Figure S2).

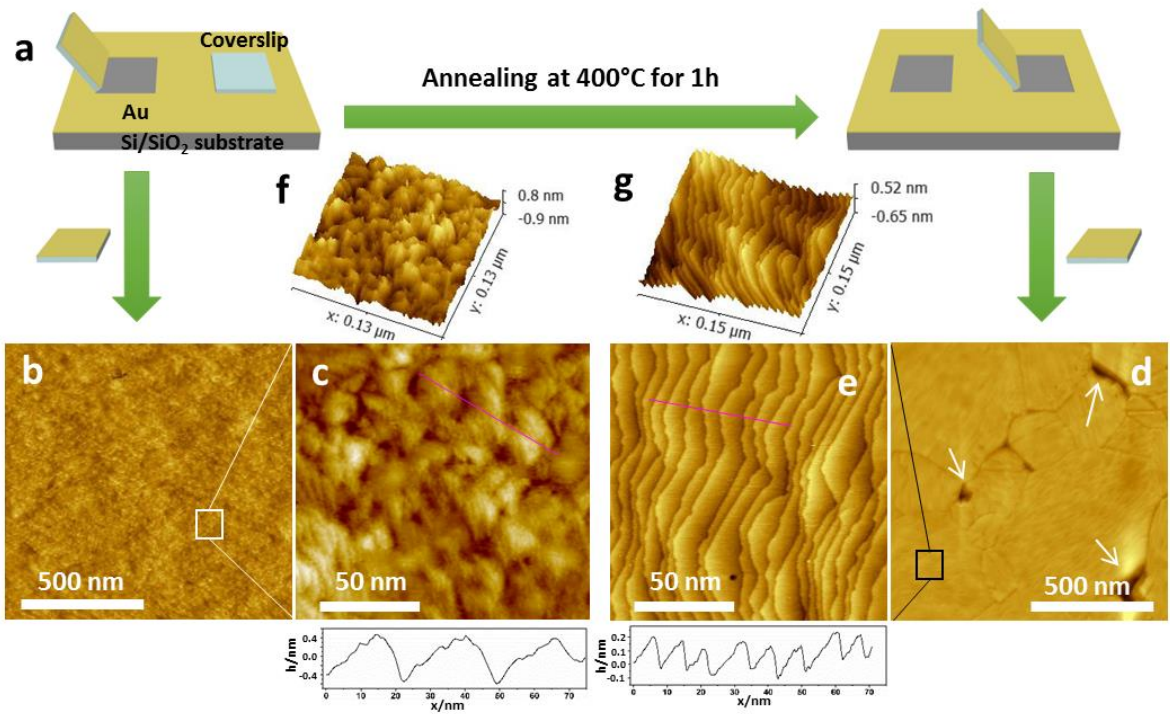

**Figure 1.** Microstructures of gold inner surfaces characterized by STM measurement. (a) Inner surfaces of gold films without annealing (Au-ref) and after annealing at 400°C for 1 h (Au-T400t1h), obtained by detaching the films from the Si/SiO$_2$ substrate. (b) STM of Au-ref showing its atomically flat surface. (c) Magnification of the area marked by a white square in (b), showing the granular nanostructures. (d) STM of Au-T400t1h, showing large grains and boundary defects. (e) Magnified image of the area marked by a black square in (d), **showing that the grains are composed of identical step structures**. (f, g) 3D images of (c, e), respectively.

Having demonstrated that annealing is essential for vicinal-like structure formation, we proceeded to investigate the effect of annealing **temperature on the step structure**. Gold samples, annealed at 100°C, 200°C and 400°C for 4 h (denoted Au-T100t4h, Au-T200t4h, and Au-T400t4h, respectively) were **prepared in a similar way**, and were characterized by STM. Well-ordered atomic steps were observed **in Au-T200t4h and Au-T400t4h,** whereas Au-T100t4h was found to be composed of roundish grains (Figure 2). These results thus already reflected efficient formation of step structures by annealing **temperatures from** around 200°C.

That the annealing temperature also affected grain size can be seen from the STM scans shown in Figures 2a−c, and from SEM imaging using the angle-selective backscatter (AsB) mode (Figure S3). Grains were found **to increase in size** with increasing annealing temperature.

The average grain size prior to annealing was 0.115 μm, whereas after annealing for 4 h at 400°C it was 0.520 μm (Table 1). **The total roughness of the samples increased** with increasing annealing **temperature, owing to** the formation of voids/cavities along grain boundaries (indicated by arrows in Figure 2b,c). At the same **time, examination of a single grain from each sample revealed** a decrease in roughness of the vicinal-like surface with increasing annealing temperature, reaching ultra-low values of 0.101 nm.

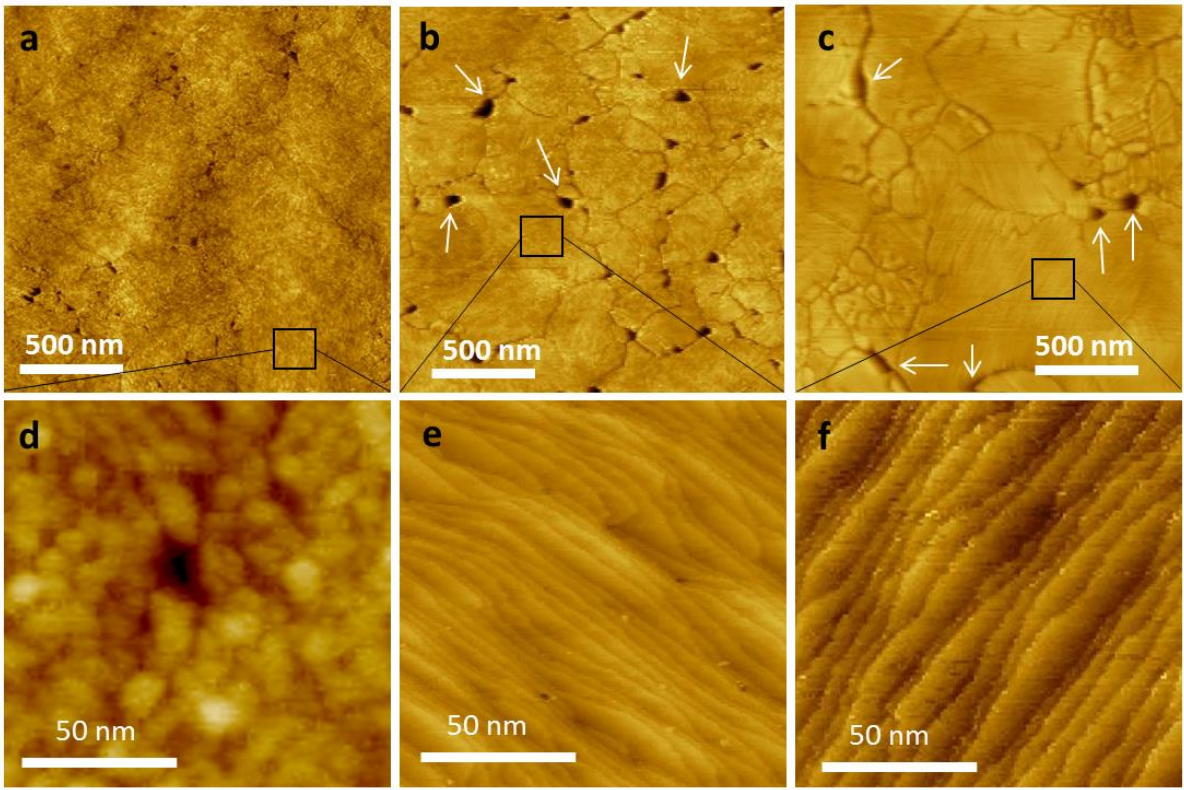

**Figure 2.** STM images of the inner surfaces of gold films after thermal treatment for 4 h. (a) Annealing at 100°C (Au-T100t4h): small pores appear. (b) Annealing at 200°C (Au-T200t4h): grains grow while pores (indicated by arrows) are clearly observed at grain boundaries. (c) Annealing at 400°C (Au-T400t4h): grains **increase in size up to ~1 μm**; grain boundaries and pores are indicated by arrows. (d), (e), and (f) Magnified images corresponding to (a) (b) and (c). Well-formed step-terrace structures are observed in Au-T200t4h and Au-T400t4h, **whereas** Au-T100t4h is composed of atomically flat nanograins.

Table 1: Grain size and roughness of gold vicinal-like surfaces at different annealing temperatures

|  | Au-ref | Au-T100t4h | Au-T200t4h | Au-T400t4h |
|---|---|---|---|---|
| **Average grain size [μm]** | 0.115 | 0.145 | 0.249 | 0.520 |
| **Total roughness [nm]** | 0.186 | 0.271 | 0.341 | 0.455 |
| **Single grain roughness [nm]** | 0.152 | 0.244 | 0.103 | 0.101 |

Additional results obtained from the STM investigation relate to the step-structure periodicity **observed** in different grains of the same sample. Step periodicity in vicinal surfaces was previously reported to correlate with the miscut angle with respect to the **low surface energy** planes[24,28]. **We postulated that** the step periodicity in our polycrystalline gold films was related to the GOA. To verify this hypothesis we carried out an *in-situ* study on **identical grains,** using electron back-scatter diffraction (EBSD) **to determine the grains' orientation angles** and STM **to determine** the average step periodicities. Gold thin film of 300-nm thickness **on our** Si/SiO$_2$ substrate was annealed at 400°C for 4 h and marked as before by a FIB with a cross on the surface to ensure that both characterization techniques **were applied to** the same area. TS was then performed as **before, allowing the inner surface of the gold to be** characterized (Figure 3).

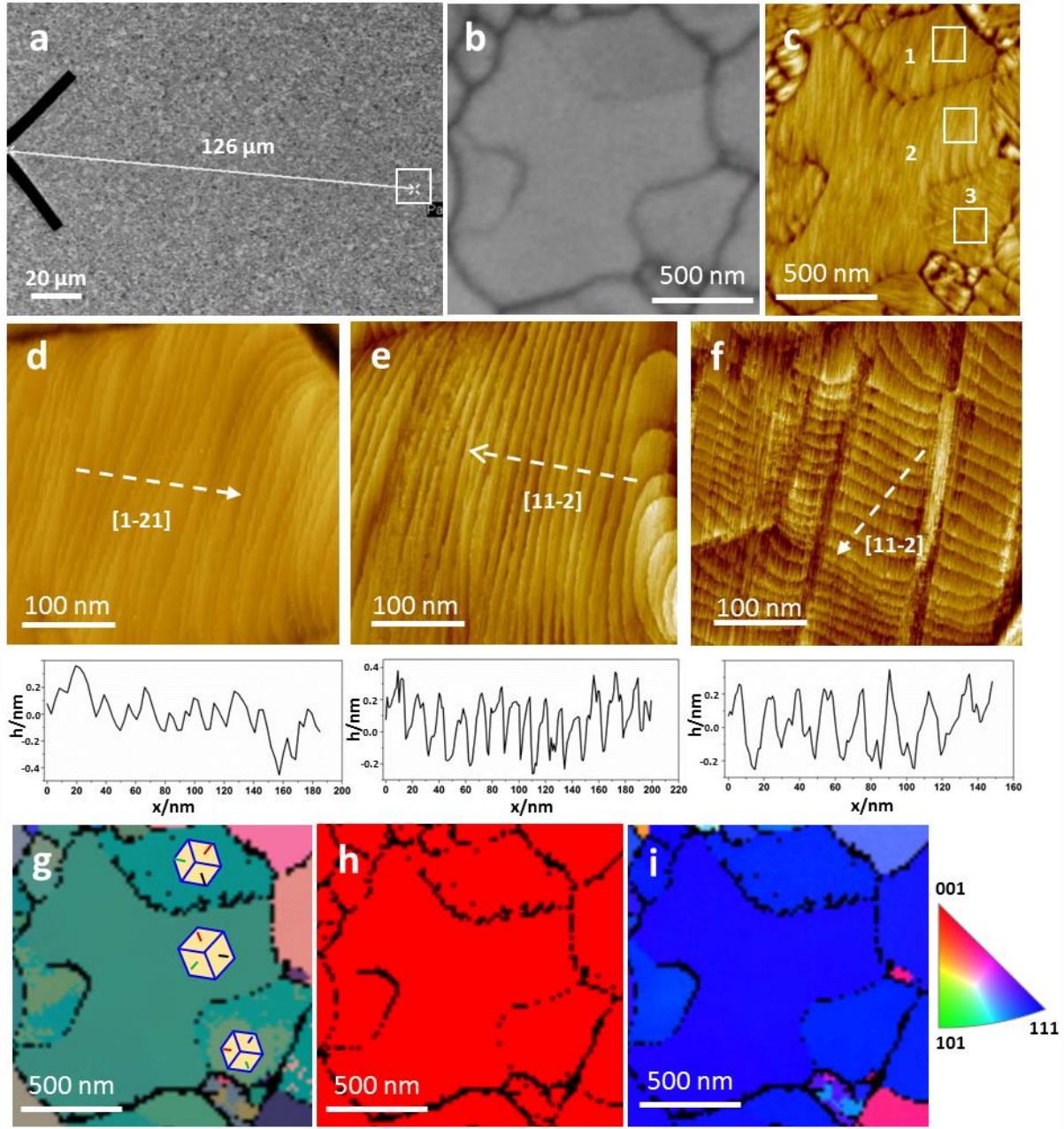

**Figure 3.** *In-situ* study of grain orientation and step periodicity. (a) AsB mode SEM image of a gold inner surface marked with a cross by FIB. (b) HRSEM zoom-in view of the area marked by a white square in (a); the square is ca. 126 μm from the center of the cross. (c) STM image of the same area as in (b). (d, e, f) Enlarged STM images of the areas marked in (c) by white squares in grains 1, 2, and 3, respectively; arrows indicate the step alignment direction, and the corresponding step **periodicities are presented under the images** by **a line** drawn along its direction. (g) Euler angle colored orientation map; crystallographic orientations of grains 1, 2, and 3 (marked in (c)) are also indicated by cubes, and the crystal coordinate system can be seen with X (red), Y (green), and Z (black). (h) Phase image, indicating the single gold phase. (i)

EBSD **Z direction inverse pole figure orientation map**. Blue indicates strong **<111>** surface normal texture.

The area marked with a white square (ca. 126 μm from the intersection of the cross) in Figure 3a **was examined** by HRSEM (Figure 3b). Several grains with obvious boundaries can be seen. The same area was also scanned by STM (Figure 3c). Grains 1, 2 and 3 are magnified and are shown in Figures 3d−f, respectively. The grains exhibit different step orientations and periodicities, while within the same grain the highly ordered steps are oriented in the same direction. To measure the step periodicity in the same crystallographic direction, **the orientation direction of the grains** should be determined first; this was done by EBSD (Figure 3g).

EBSD demonstrated that the sample was a single gold phase (Figure 3h), further confirming the sample's purity. Figure 3i is the **Z direction inverse pole figure orientation map**. The blue color indicates that the Z direction is strongly parallel to the {111} direction of the gold **grains**, and shows that that the orientation angles of the {111} planes are **small relative** to the substrate. However, the grains demonstrate different in-plane orientations, which were confirmed by the Euler angle coloring orientation map of Figure 3g. The crystallographic orientations of grains 1, 2 and 3 are indicated by cubes. Normally the steps were directed **to one direction of the <112> family** in each grain, as determined from the EBSD results (Figure S5) and indicated by arrows in Figures 3d−f. The <112> is the close-packed **direction and the 'classical' vicinal surfaces obtained by miscutting usually grow in this direction**.[28,29] In our vicinal-like surfaces the terraces are also aligned in the <112> direction. Further analysis, combined with our STM results, confirmed that the step structures are aligned along the <112> direction. The average step periodicities were therefore determined by **drawing lines** along the <112> direction (e.g., images below Figure 3d−f, respectively). The average step periodicities of grains 1, 2 and 3 were 12.75 ± 0.27 nm, 12.54 ± 0.23 nm, and 16.26 ±0.73 nm, respectively.

The accurate orientation angles (α) of gold **(111) planes relative** to the substrate can be calculated according to the equation: $\cos\alpha = \frac{h+k+l}{\sqrt{3}}$, where $h$, $k$, and $l$ are the normalized Miller index values of the crystal (HKL) **plane, defined as being parallel to the substrate surface X−Y (see Figure S6) *and hence can be obtained*** by $h = \sin\varphi_2 \sin\emptyset$, $k = \cos\varphi_2 \sin\emptyset$, $l = \cos\emptyset$. Angles $\emptyset$ and $\varphi2$ are the Euler angles collected by EBSD (Table S1). The **calculation yields** orientation angles of 5.10 ± 0.05 degrees, 4.54 ± 0.05 degrees and 4.01 ± 0.03 degrees for grains 1, 2 and 3, respectively. Two additional grains of the same sample were analyzed in a similar way (Figure S7, grains 4 and 5), and the correlation between **the grains'** orientation angles and the step periodicities is depicted in Figure 4. We found that step periodicity is correlated with the **grains' orientation angles**, meaning that a smaller misorientation between the (111) gold interface and the substrate will result in larger step periodicity, as shown in Figure 4.

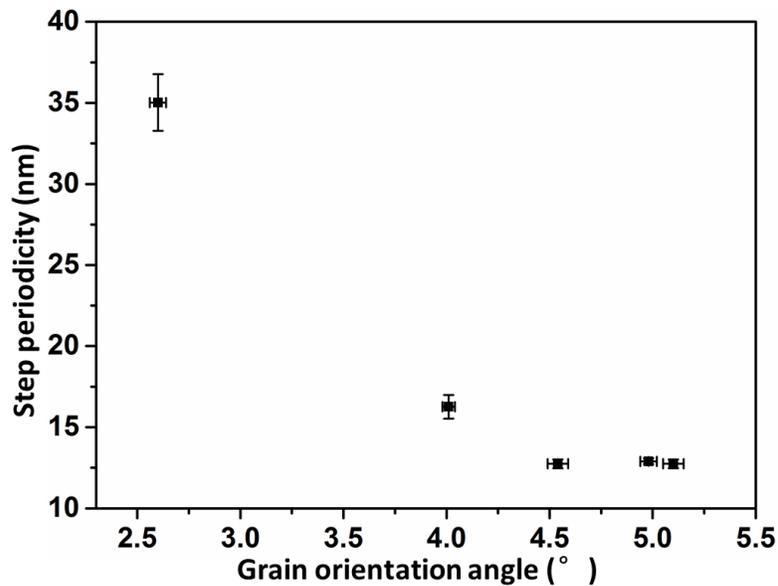

**Figure 4.** Correlation **between a between grains' orientation angles** and step periodicities. With increasing grain orientation, the step periodicity decreases.

The formation of vicinal-like structures with the lowest energy (111) terraces is driven by minimization of the surface **energy, which includes** step formation and step-step interaction energy [28]. Formation of (111) terraces is controlled **by diffusion-driven** transport of gold atoms from the inner gold/substrate interface to the bulk and the outer surface of the film. Thermal annealing at 400°C for 4 h provides substantial bulk diffusion across the film, such that the resulting step **structure can be** expected **to approach** the equilibrium one. Relaxation of atoms around the step edge is a dipolar elastic distortion, and these strain fields give rise to long-range $(1/d^2)$ step-step interactions that are predominantly repulsive. [30, 31]. Such repulsive step-step interaction may result **in increasing** periodicity during annealing, **as indeed** observed in our experimental investigation of the vicinal-like film/substrate interfaces. After T400t4h annealing the periodicity reached a certain minimum saturation value (about 12 nm) for misorientation angles above 4.5 degrees, possibly indicative of its approach to the equilibrium state of the interface.

**It is therefore reasonable to** conclude **that formation of a vicinal-like surface with periodic step structure during thermal annealing** is driven by minimization of the surface energy. Orientation of **grains relative to** the inner gold/substrate interface determines the step periodicity, while thermal annealing may result in formation of a near-equilibrium state of the interface with **a saturated step periodicity value**.

In **this** *in-situ* study, self-ordered vicinal-like atomic structures were obtained by annealing of thin films of gold at temperatures of 200° to 400°C. Annealing at temperatures ≥200°C efficiently promoted the formation of vicinal-like structures. Gold grains near the inner **surface** exhibited the [111] **direction close to the normal to the substrate and presented random in-plane orientation**. Furthermore, the step periodicity depended on misorientation of the grain/substrate angle. Smaller misorientation resulted in a larger average step periodicity, similar to that seen in regular vicinal surfaces of gold single crystals. The formation of vicinal-like *structures* **could be attributed** to the thermally activated surface reconstruction driven by minimization of the surface energy. This work contributes to the understanding of the formation of atomic step structures at the gold/substrate interface, which will be helpful in the use of vicinal-like surfaces as templates for the growthe of regularly spaced nanostructures. It **also offers** a method for the *in-situ* investigation of orientation and step periodicity **in identical grains, and this method** can then be utilized for **further explorations of** vicinal surfaces.

**EXPERIMENTAL**

*Sample Preparation.* Samples were prepared according to our previous method[27] and as shown in Figure S1. We obtained thin (100-nm or 300-nm) polycrystalline Au films by evaporating thin films of gold (99.99% pure, Sigma Aldrich) onto a clean (001) silicon **wafer possessing** an atomically flat layer of thermally grown silica (ca. 100 nm thick). Evaporation was performed in an E-beam-equipped AircoTemescal FC-1800 evaporating system with a deposition rate of 2.7 Å/s **under an ultra-high** vacuum of $10^{-7}$ torr. Gold films on a $SiO_2$ substrate were annealed in a Carbolite 201 **oven (Eurotherm) in an air atmosphere** at different temperatures (100°C, 200°C and 400°C) for 1 h or 4 h. On each of the samples to be examined by both EBSD and STM we made a cross mark thicker than the film on its gold surfaces, utilizing a FIB to ensure that these characterization techniques would be applied in all of them at identical positions. A coverslip (3 mm × 8 mm), pre-cleaned in piranha solution, was attached to **the** **upper** gold surface by means of a UV-curable adhesive (Norland, NOA 61) or an electrically conductive adhesive (Norland, NCA 130), exposed to UV for 0.5 h, and then heat-cured at 50°C for 5 h in an oven. Using a scalpel blade we then stripped the glass off the substrate. Then, by stripping the coverslip together with the gold thin film from the substrate **the lower** surface of the gold **was exposed,** and was subjected to STM to characterize its atomic structure (denoted as Au-ref, Figure 1a−c). A similar sample (substrate and gold thin film) was annealed at 400°C for 1 h, and then similar TS was performed, followed by STM (denoted as Au-T400t1h, Figure 1d, e).

*Characterization.* We checked the purity of **the lower** gold surfaces by X-ray photoelectron spectroscopy (XPS), using a Thermo VG SIGMA probe. STM was performed in an Omicron MATRIX Control System with a vacuum of $4 \times 10^{-11}$ torr. The EBSD system (**AZtec[?AZtecHKL**]) used **for this** *in-situ* study was attached to an FEI Quanta 200 scanning electron microscope (SEM) equipped with a field emission gun. **The sample** was tilted to 70° from the horizontal towards the EBSD detector. Measurements were carried out at a working distance of 11 mm and an accelerating voltage of 20 kV. To determine grain orientations, we scanned an area of 2 μm × 2 μm at step intervals of 20 nm. SEM was performed with a Zeiss ultra-Plus HR-SEM, using the AsB mode to observe the grain structure. The working voltage was 10 kV and the working distance was 2.8 mm.

### ASSOCIATED CONTENT

#### Supporting Information

The Supporting Information is available free of charge.

gold film sample preparation, XPS results, SEM images in AsB mode, determination of the Miller Index of the step orientation direction, Euler angles obtained by EBSD measurements (PDF).

**AUTHOR INFORMATION**

**Corresponding Author**

*E-mail: bpokroy@tx.technion.ac.il

**Notes**

**The authors declare no competing financial interest.**

**ACKNOWLEDGMENTS**

The research leading to these results received funding from the European Research Council under the European Union's Seventh Framework Program (FP/2007-2013)/ERC Grant Agreement (no. 336077). L.C. acknowledges financial support by a Fellowship from the Council for Higher Education in Israel and the Technion Fund for Cooperation with Far-East Universities. M.K.K. acknowledges with thanks the financial support from the Israeli Ministry of Science, Technology and Space.

**REFERENCES**

(1)     Sun, L.; Diaz-Fernandez, Y. A.; Gschneidtner, T. A.; Westerlund, F.; Lara-Avila, S.; Moth-Poulsen, K., Single-molecule electronics: from chemical design to functional devices. *Chemical Society Reviews* 2014, 43, (21), 7378-7411.

(2)     Wang, X.; Aroonyadet, N.; Zhang, Y.; Mecklenburg, M.; Fang, X.; Chen, H.; Goo, E.; Zhou, C., Aligned Epitaxial SnO2 Nanowires on Sapphire: Growth and Device Applications. *Nano Letters* 2014, 14, (6), 3014-3022.

(3)     Yao, J.; Yan, H.; Lieber, C. M., A nanoscale combing technique for the large-scale assembly of highly aligned nanowires. *Nat Nano* 2013, 8, (5), 329-335.

(4)     Lu, W.; Lieber, C. M., Nanoelectronics from the bottom up. *Nat Mater* 2007, 6, (11), 841-850.

(5)     Hu, J.; Wu, R., Giant Magnetic Anisotropy of Transition-Metal Dimers on Defected Graphene. *Nano Letters* 2014, 14, (4), 1853-1858.

(6)     Loth, S.; Baumann, S.; Lutz, C. P.; Eigler, D. M.; Heinrich, A. J., Bistability in Atomic-Scale Antiferromagnets. *Science* 2012, 335, (6065), 196-199.

(7)     Emboras, A.; Niegemann, J.; Ma, P.; Haffner, C.; Pedersen, A.; Luisier, M.; Hafner, C.; Schimmel, T.; Leuthold, J., Atomic Scale Plasmonic Switch. *Nano Letters* 2015.

(8)     Shomroni, I.; Rosenblum, S.; Lovsky, Y.; Bechler, O.; Guendelman, G.; Dayan, B., All-optical routing of single photons by a one-atom switch controlled by a single photon. *Science* 2014, 345, (6199), 903-906.

(9)     Pathem, B. K.; Claridge, S. A.; Zheng, Y. B.; Weiss, P. S., Molecular Switches and Motors on Surfaces. *Annual Review of Physical Chemistry* 2013, 64, (1), 605-630.

(10)     Klappenberger, F., Two-dimensional functional molecular nanoarchitectures – Complementary investigations with scanning tunneling microscopy and X-ray spectroscopy. *Progress in Surface Science* 2014, 89, (1), 1-55.

(11)     Barth, J. V.; Costantini, G.; Kern, K., Engineering atomic and molecular nanostructures at surfaces. *Nature* 2005, 437, (7059), 671-679.

(12)     Shang, J.; Wang, Y.; Chen, M.; Dai, J.; Zhou, X.; Kuttner, J.; Hilt, G.; Shao, X.; Gottfried, J. M.; Wu, K., Assembling molecular Sierpiński triangle fractals. *Nat Chem* 2015, 7, (5), 389-393.

(13)     Jia, C.; Wang, J.; Yao, C.; Cao, Y.; Zhong, Y.; Liu, Z.; Liu, Z.; Guo, X., Conductance Switching and Mechanisms in Single-Molecule Junctions. *Angewandte Chemie International Edition* 2013, 52, (33), 8666-8670.

(14)     Kwiat, M.; Cohen, S.; Pevzner, A.; Patolsky, F., Large-scale ordered 1D-nanomaterials arrays: Assembly or not? *Nano Today* 2013, 8, (6), 677-694.

(15)     Tao, F.; Dag, S.; Wang, L.-W.; Liu, Z.; Butcher, D. R.; Bluhm, H.; Salmeron, M.; Somorjai, G. A., Break-Up of Stepped Platinum Catalyst Surfaces by High CO Coverage. *Science* 2010, 327, (5967), 850-853.

(16)     Saywell, A.; Schwarz, J.; Hecht, S.; Grill, L., Polymerization on Stepped Surfaces: Alignment of Polymers and Identification of Catalytic Sites. *Angewandte Chemie* 2012, 124, (21), 5186-5190.

(17)     Battaglini, N.; Repain, V.; Lang, P.; Horowitz, G.; Rousset, S., Self-assembly of an octanethiol monolayer on a gold-stepped surface. *Langmuir* 2008, 24, (5), 2042-2050.

(18)    Lee, S.; Bae, S.-S.; Medeiros-Ribeiro, G.; Blackstock, J. J.; Kim, S.; Stewart, D. R.; Ragan, R., Scanning tunneling microscopy of template-stripped Au surfaces and highly ordered self-assembled monolayers. *Langmuir* 2008, 24, (12), 5984-5987.

(19)    Tsivion, D.; Schvartzman, M.; Popovitz-Biro, R.; von Huth, P.; Joselevich, E., Guided Growth of Millimeter-Long Horizontal Nanowires with Controlled Orientations. *Science* 2011, 333, (6045), 1003-1007.

(20)    Goren-Ruck, L.; Tsivion, D.; Schvartzman, M.; Popovitz-Biro, R.; Joselevich, E., Guided Growth of Horizontal GaN Nanowires on Quartz and Their Transfer to Other Substrates. *Acs Nano* 2014, 8, (3), 2838-2847.

(21)    Arora, S. K.; O'Dowd, B. J.; Ballesteros, B.; Gambardella, P.; Shvets, I. V., Magnetic properties of planar nanowire arrays of Co fabricated on oxidized step-bunched silicon templates. *Nanotechnology* 2012, 23, (23), 235702.

(22)    Xiao, W.; Ruffieux, P.; Aït-Mansour, K.; Gröning, O.; Palotas, K.; Hofer, W. A.; Gröning, P.; Fasel, R., Formation of a Regular Fullerene Nanochain Lattice. *The Journal of Physical Chemistry B* 2006, 110, (43), 21394-21398.

(23)    Tsivion, D.; Schvartzman, M.; Popovitz-Biro, R.; Joselevich, E., Guided Growth of Horizontal ZnO Nanowires with Controlled Orientations on Flat and Faceted Sapphire Surfaces. *Acs Nano* 2012, 6, (7), 6433-6445.

(24)    Butashin, A. V.; Vlasov, V. P.; Kanevskii, V. M.; Muslimov, A. E.; Fedorov, V. A., Specific features of the formation of terrace-step nanostructures on the (0001) surface of sapphire crystals. *Crystallogr Rep* 2012, 57, (6), 824-830.

(25)    Borukhin, S.; Saguy, C.; Koifman, M.; Pokroy, B., Self-Ordered Vicinal-Surface-Like Nanosteps at the Thin Metal-Film/Substrate Interface. *The Journal of Physical Chemistry C* 2012, 116, (22), 12149-12155.

(26)    Ragan, R.; Ohlberg, D.; Blackstock, J. J.; Kim, S.; Williams, R. S., Atomic surface structure of UHV-prepared template- stripped platinum and single-crystal platinum(111). *J Phys Chem B* 2004, 108, (52), 20187-20192.

(27)    Borukhin, S.; Saguy, C.; Koifman, M.; Pokroy, B., Self-Ordered Vicinal-Surface-Like Nanosteps at the Thin Metal-Film/Substrate Interface. *J Phys Chem C* 2012, 116, (22), 12149-12155.

(28)    Rousset, S.; Repain, V.; Baudot, G.; Garreau, Y.; Lecoeur, J., Self-ordering of Au(111) vicinal surfaces and application to nanostructure organized growth. *J Phys-Condens Mat* 2003, 15, (47), S3363-S3392.

(29)    Ortega, J.; Mugarza, A.; Repain, V.; Rousset, S.; Pérez-Dieste, V.; Mascaraque, A., One-dimensional versus two-dimensional surface states on stepped Au (111). *Physical review B* 2002, 65, (16), 165413.

(30) J. F. Annett, First-Principles Calculation of Surface Step Energies and Interactions, In: Kumar V., Sengupta S., Raj B. (eds) Frontiers in Materials Modelling and Design. Springer, Berlin, Heidelberg., 271.

(31) B. Houchmandzadeh and C. Misbah, Elastic Interaction Between Modulated Steps on  a Vicinal Surface, J. Phys. I France 5, 685 (1995).